\documentclass[prl,aps,graphics,floats,showpacs,preprint]{revtex4}
\usepackage{graphicx}
\usepackage{amssymb}

\usepackage{amsmath}

\begin{document}

\draft

\title{A universal scaling law of exchange bias training effect}

\author{Z. Shi$^{\mathrm{1}}$, S. M. Zhou$^{\mathrm{1}}$, and S. Mangin$^{\mathrm{2}}$}

\address{$^{\mathrm{1}}$Department of Physics, Tongji University, Shanghai 200092, China}

\address{$^{\mathrm{2}}$Nancy University, Inst Jean Lamour, F-54506 Vandoeuvre Les Nancy, France}

\date{\today}

\begin{abstract} Exchange bias training effect in ferromagnetic/antiferromagnetic bilayers is investigated. In some systems the evolution of the exchange bias field $H_E$ with the number of cycle $n$ cannot be fitted by the empirical $1/\sqrt{n}$ function. A unified expression is derived from a discretized Landau-Khalatnikov equation in the framework of the thermodynamics model which is proposed by Ch.\ Binek. This generalized model describes well training effect independent of the magnetization reversal mechanism in the ferromagnetic layers. \\


\end{abstract}

\vspace{0.5 cm}

\maketitle

\newpage
\indent
Exchange bias (EB) effect has been investigated extensively since its first discovery in CoO/Co nanoparticles~\cite{1,2,3,4,5}. After the EB
is established in antiferromagnetic(AFM)/ferromagnetic(FM) bilayers by (magnetic) field cooling from high temperature to below the N\'{e}el temperature of the AFM layer or by magnetic field during film deposition, a shift of the hysteresis loop along the magnetic cooling field axis with an exchange field ($H_E$) is observed, companied by an enhancement of the coercivity ($H_C$). When cycling the bilayer through consecutive hysteresis loops, $H_E$ and $H_C$ often decrease monotonically with increasing cycling number $n$~\cite{6}. This so-called training effect has attracted extensive investigation because of its importance in both basic research and applications of spintronic devices~\cite{6,7,8,9,10,11,12,13,14,15,16,17,18,19,20,21}.\\
\indent During the training effect, large changes in $H_E$ and $H_C$ are often observed between $n=1$ and $n=2$, which is thought to be strongly related to transition of AFM spins between easy axis~\cite{11}. For $n\geq 2$, $H_E$ decreases gradually with increasing $n$, which can be described by the empirical formula~\cite{6,8,10,15,19}.
\begin{equation}
H_{E}(n)=H_{E}(\infty)+\frac{k}{\sqrt{n}}
\label{Eq1}
\end{equation}
where $k$ is a parameter depending on the physics properties of the AFM layer. The empirical law was observed first for Co/CoO, NiFe/NiFeMn, and $\mathrm{NiFe/Cr_{2}O_{3}}$ systems~\cite{6}. The EB training effect was considered in the framework of nonequilibrium thermodynamics by Ch.\ Binek, where the free energy of the system controls the relaxation process of the AFM spins towards the equilibrium state. With a discretized Landau-Khalatnikov equation, the following equation is obtained~\cite{10}.
\begin{equation}
H_{E}(n)-H_{E}(n+1)=\gamma(H_{E}(n)-H_{E}(\infty))^3
\label{Eq2}
\end{equation}
where $\gamma$ is the characteristic decay rate of the training behavior. Since Eq.~\ref{Eq1} does not hold for some EB systems, such as NiFe/IrMn bilayers and Pt/Co/Pt/IrMn multilayers~\cite{16,17}, several approaches have been proposed. For example, higher orders in the free energy were considered~\cite{15}. Alternatively, based on a thermal fluctuation model, a power law function on the cycle number was used to fit the training effect of $H_E$~\cite{8}. Moreover, the training effect was suggested to be strongly related to the evolution of the AFM spin disorder and the training effect of $H_E$ can be expressed by an exponential function~\cite{14,16}. In this work, we will derive a unified equation to describe the variation of $H_E$ with $n$ based on the theoretical model proposed by Binek, which can describe the EB training effect in FM/AFM bilayers with various FM magnetization reversal mechanisms.\\
\indent
Two sets of samples were prepared by DC magnetron sputtering at ambient temperature: glass/Cu(20 nm)/Ni$_{80}$Fe$_{20}$(=NiFe)(6 nm)/Fe$_{50}$Mn$_{50}$(=FeMn)(11 nm)/Au(10 nm) and glass/Cu(20 nm)/Fe$_{x}$Cr$_{1-x}$/Ir$_{25}$Mn$_{75}$(=IrMn)(3 nm)/Cu(10 nm) with varying $x$. The base pressure of the system was $1.8\times10^{-5}$ Pa and the working Argon pressure was 0.5 Pa during deposition. The deposition rates of the constituent layers were in the order of 0.1 nm/second. A magnetic field of 170 Oe was applied in the film plane during the film deposition to set the longitudinal exchange bias. In-plane magnetization loops were measured at room temperature by a Lakeshore vector vibrating sample magnetometer (VVSM). During training effect measurements, consecutive hysteresis loops were measured after deposition without break. All measurements were performed at room temperature. \\
\indent Figure~\ref{Fig1} shows typical hysteresis loops at room temperature along the easy axis of NiFe/FeMn, Fe$_{0.51}$Cr$_{0.49}$/IrMn, and Fe$_{0.36}$Cr$_{0.64}$/IrMn bilayers. For the three samples, $H_E$ at $n=1$ equals 54.7, 14.5, and 24.5 (Oe) and is reduced by 28.9, 14.2, and 14.65 (Oe) after measurements of 160, 80, and 80 consecutive hysteresis loops, respectively. Large training effect can be found in all samples. Moreover, for NiFe/FeMn and Fe$_{0.51}$Cr$_{0.49}$/IrMn bilayers, the coercive field of the descending branch changes strongly while that of the ascending branch is hardly modified, resulting in a decrease of the coercivity. For Fe$_{0.36}$Cr$_{0.64}$/IrMn bilayers, however, both branches shift towards the positive magnetic field direction and the coercivity almost does not change, i.e., from 28 to 25.5 (Oe) after 80 cycles (Fig.~\ref{Fig1}(c)), as already observed for the perpendicularly magnetized Pt/Co/Pt/IrMn bilayers~\cite{17}.\\
\indent As shown in Fig.~\ref{Fig2}, the measured $H_{E}(n)$ in NiFe/FeMn, Fe$_{0.51}$Cr$_{0.49}$/IrMn, and Fe$_{0.36}$Cr$_{0.64}$/IrMn bilayers are fitted by using Eq.~\ref{Eq2}. One can easily find that the experimental results can be fitted by Eq.~\ref{Eq2}. Therefore, the EB training effect in these three samples can be described a discretized Landau-Khalatnikov equation. In principle, $\gamma$ and $H_{E}(\infty)$ can be fitted by using Eq.~\ref{Eq2}. Since $H_{E}(n)-H_{E}(n+1)$ does not change much with $n$ except for $n=1$, however, the fitted values of these parameters are not rigorous. \\
\indent In order to examine whether the measured  $H_{E}(n)$ can be described by the empirical law in Eq.~\ref{Eq1}, $H_E$ for NiFe/FeMn and Fe$_{0.51}$Cr$_{0.49}$/IrMn bilayers is plotted as a function of $1/\sqrt{n}$ in Fig.~\ref{Fig3}. For different systems, 80 to 160 hysteresis loop cycles were measured to get an overall feature of the training effect. Only few tens of magnetization loops are usually measured~\cite{6,7,9,10,19,20,21}, which is not \emph{adequate} to inquire into the analytical expression of the training effect. In comparison, the data of epitaxially grown Fe/CoO bilayers are also given~\cite{16}. Here, all samples exhibit the conventional training effect in which $H_E$ decreases sharply between $n=1$ and $n=2$ and then changes slowly for large $n$~\cite{7}. It is found that the measured data cannot be described very well by the empirical linear function of $1/\sqrt{n}$ even for $n\geq2$.
 Based on these observations, we re-derived the dependence of $H_E$ on the cycle number. Starting with Eq.~\ref{Eq2}~\cite{10}, the differential of $H_E$ between neighboring consecutive hysteresis loops $H_E(n)-H_E(n+1)$ is proportional to $(H_E{(n)}-H_E{(\infty)})^3$ when $H_E$ changes \emph{slowly} with $n$. Accordingly, one has the following differential equation,
\begin{equation}
\frac{\mathrm{d}H_E(n)}{\mathrm{d}n}=-\gamma(H_E{(n)}-H_E{(\infty)})^3
\label{Eq4}
\end{equation}
where $\mathrm{d}n=1$ can then be obtained. Since $H_E$ often changes slowly for $n\geq 2$, the analytical solution of the $n$-dependent $H_E(n)$ can be achieved as follows.
\begin{equation}
H_E(n)=H_E(\infty)+\frac{1}{\sqrt{2\gamma}\sqrt{n+n_0-2}}
\label{Eq5}
\end{equation}
where $n_0={H_E(2)-H_E(\infty) \over 2(H_E(2)-H_E(3))}$, strongly reflecting the fraction of the $H_E$ change between $n=2$ and 3 in the entire training effect for $n\geq 2$. Similar parameters have been defined before~\cite{23a}. If $n_0=2.0$ in above equation, Eq.~\ref{Eq5} turns to the conventional empirical law, i.e., $H_E(n)-H_E(\infty)=\frac{1}{\sqrt{2\gamma n}}$. $H_E(\infty)$, $\gamma$, and $n_0$ can be fitted with the measured $H_E(n)$. Figures~\ref{Fig3}(d)-~\ref{Fig3}(f) show that the fitting results and experiment data are in good agreement for NiFe/FeMn, Fe$_{0.51}$Cr$_{0.49}$/IrMn, and Fe/CoO bilayers. The value of $n_0$ is fitted to be 6.6, 5.8, and 0.97, and $\gamma$ to be $3.1\times10^{-4}$, $5.5\times10^{-4}$, and $7.0\times10^{-4}$ $(\mathrm{Oe}^{-2})$, respectively. \\
\indent To further confirm its universal validity, Eq.~\ref{Eq5} is also used to analyze the training effect for Fe$_{0.36}$Cr$_{0.64}$/IrMn bilayers and perpendicularly exchange-biased Pt/Co/Pt/IrMn multilayers~\cite{18}. Here, both samples exhibit the anomalous training effect~\cite{7}.
Note that $H_E(n)$ of neither sample can be fitted with the $1/\sqrt{n}$ function, as shown in Figs.~\ref{Fig4}(a) $\&$~\ref{Fig4}(b).
Actually, for FeCr/IrMn bilayers the measured data \emph{at any n regime} cannot be fitted by $1/\sqrt{n}$. However, the measured data can be described by Eq.~\ref{Eq5} very well as shown in Figs.~\ref{Fig4}(c) $\&$~\ref{Fig4}(d). The $n_0$ is found to be 10.3 and 6.5 and $\gamma$ is $2.3\times10^{-4}$ and $1.3\times10^{-6}$ $\mathrm{(Oe^{-2})}$ for Fe/CoO and Pt/Co/Pt/IrMn multilayers, respectively. it is therefore evident that the model based on the AFM spin relaxation process holds for training effects in various FM/AFM bilayers with either the conventional training effect or the anomalous training effect.\\
\indent When the magnetization reversal is dominated by domain wall depinning, such as in Fe$_{0.36}$Cr$_{0.64}$/IrMn and Pt/Co/Pt/IrMn systems, $H_E$ shows the so-called anomalous training effect in which $H_E(1)-H_E(2)$ is small, i.e. the first-jump vanishes~\cite{11}. Meanwhile, when the magnetization reversal is dominated by magnetization coherent rotation, such as in NiFe/FeMn, Fe$_{0.51}$Cr$_{0.49}$/IrMn, and Fe/CoO bilayers, $H_E$ exhibits conventional training effect in which the first-jump appears~\cite{11,16,17,18,22a}. Therefore, the magnetization reversal mechanism plays an important role in the training effect. Moreover, when the evolution of $H_E(n)$ deviates seriously from the $1/\sqrt{n}$ linear dependence, such as for Pt/Co/Pt/IrMn, FeCr/IrMn, and NiFe/FeMn systems, a large $n_0$ is acquired. Otherwise, a small $n_0$ is obtained, such as for the Fe/CoO bilayer. Furthermore, according to $n_0={H_E(2)-H_E(\infty) \over 2(H_E(2)-H_E(3))}$ and Eq.~\ref{Eq2}, one has $H_E(2)-H_E(\infty)={1 \over \sqrt{2n_0\gamma}}$, that is to say, the $n_0\gamma$ product determines the training effect for $n\geq 2$, i.e., $H_E(2)-H_E(\infty)$. Finally, with the values of the parameters $\gamma$ and $n_0$, the $n$ dependence of $H_E(n)$ can be identified.\\
\indent In summary, for NiFe/FeMn, Fe$_{0.36}$Cr$_{0.64}$/IrMn, and Fe$_{0.51}$Cr$_{0.49}$/IrM bilayers the  magnetization reversal process is dominated by domain rotation, the domain wall depinning, and a combination of both modes, respectively. Depending on the reversal process, the evolution of $H_E(n)$ may be close to or seriously deviate from the empirical $1/\sqrt{n}$ function. Based on the thermodynamics model proposed by Ch.\ Binek~\cite{10}, a unified analytical function of $1/\sqrt{n+n_0-2}$ is derived for $n\geq2$ and is verified to be applicable to the thermally activated EB training for various FM/AFM bilayers.\\
\indent  This work was supported by the National Natural Science Foundation of China under grant Nos. 10974032 and 51171129, the National Basic Research Program of China under grant No. 2009CB929201.\\


\begin{figure}[p]
\begin{center}

FIGURE CAPTIONS
\end{center}

\flushleft \indent Figure 1 Typical hysteresis loops of NiFe/FeMn (a), Fe$_{0.51}$Cr$_{0.49}$/IrMn (b), and Fe$_{0.36}$Cr$_{0.64}$/IrMn (c) bilayers. The inset numbers refer to the cycle numbers of hysteresis loops.\\

\indent Figure 2
The curve of the measured $H_E(n)-H_E(n+1)$ versus $H_E(n)$ for NiFe/FeMn (a), Fe$_{0.51}$Cr$_{0.49}$/IrMn (b), and Fe$_{0.36}$Cr$_{0.64}$/IrMn (c) bilayers. The solid lines refer to the fitted results based on Eq.~\ref{Eq2}.\\

\indent Figure 3 The $n$ dependence of the measured $H_E$ (open symbols) for NiFe/FeMn (a,d), Fe$_{0.51}$Cr$_{0.49}$/IrMn (b,e), and epitaxial Fe/CoO~\cite{17} (c,f) bilayers. The measured data were fitted using a linear function of $1/\sqrt{n}$ (the left column) and $1/\sqrt{n+n_0-2}$ (the right column), respectively. The solid lines refer to the fitted results based on Eq.~\ref{Eq5}.\\

\indent Figure 4
The $n$ dependence of  the measured $H_E$ for Fe$_{0.36}$Cr$_{0.64}$/IrMn bilayers (a,c)~\cite{16} and Pt/Co/Pt/IrMn multilayer (b,d). The measured data were fitted using a linear function of $1/\sqrt{n}$ (the left column) and $1/\sqrt{n+n_0-2}$ (the right column), respectively. The solid lines refer to the fitted results based on Eq.~\ref{Eq5}.\\

\end{figure}

\newpage
\begin{figure}[p]
\begin{center}
\resizebox*{6 in}{!}{\includegraphics*{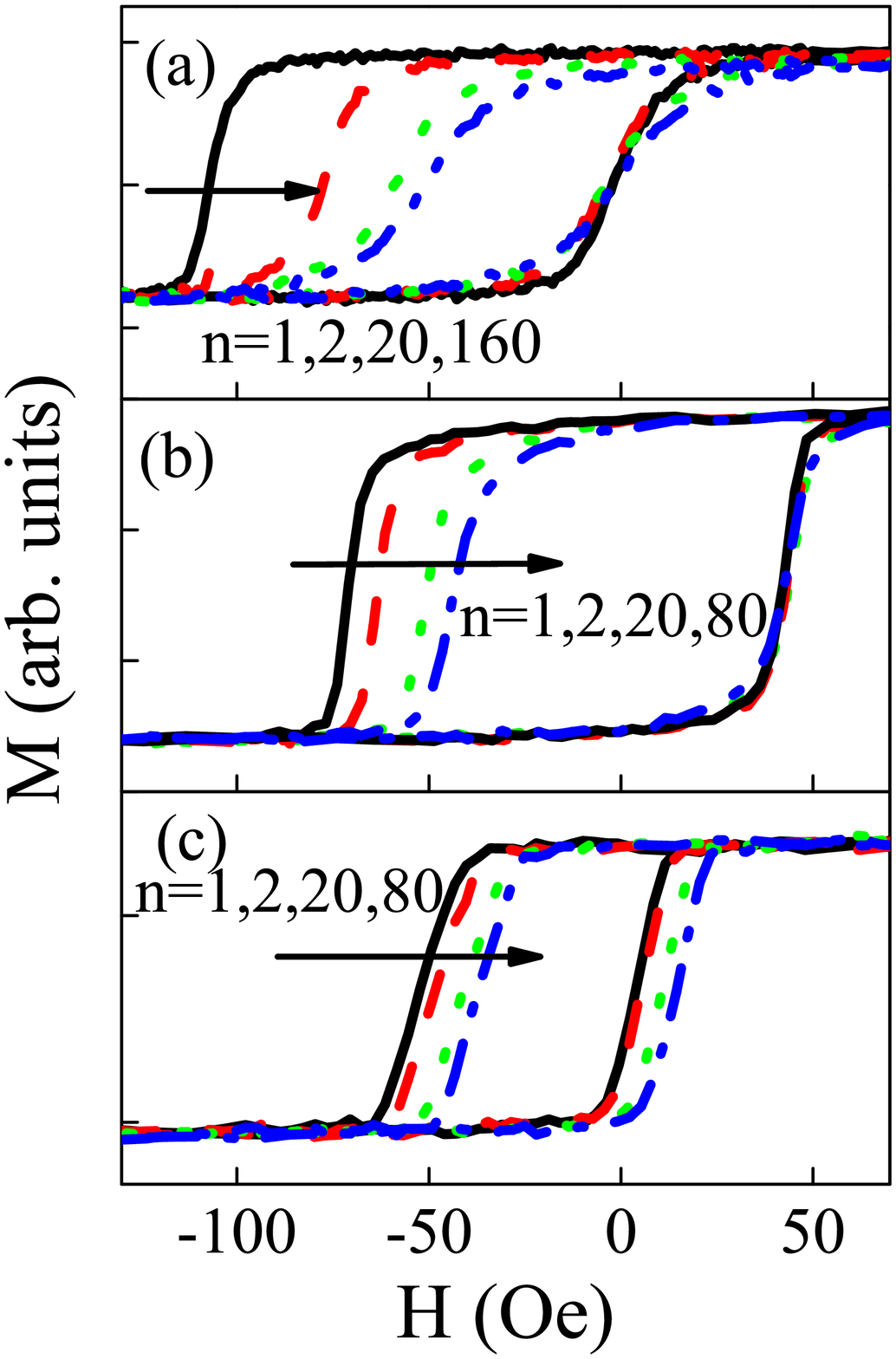}} \caption{}
\label{Fig1}
\end{center}
\end{figure}

\begin{figure}[p]
\begin{center}
\resizebox*{6 in}{!}{\includegraphics*{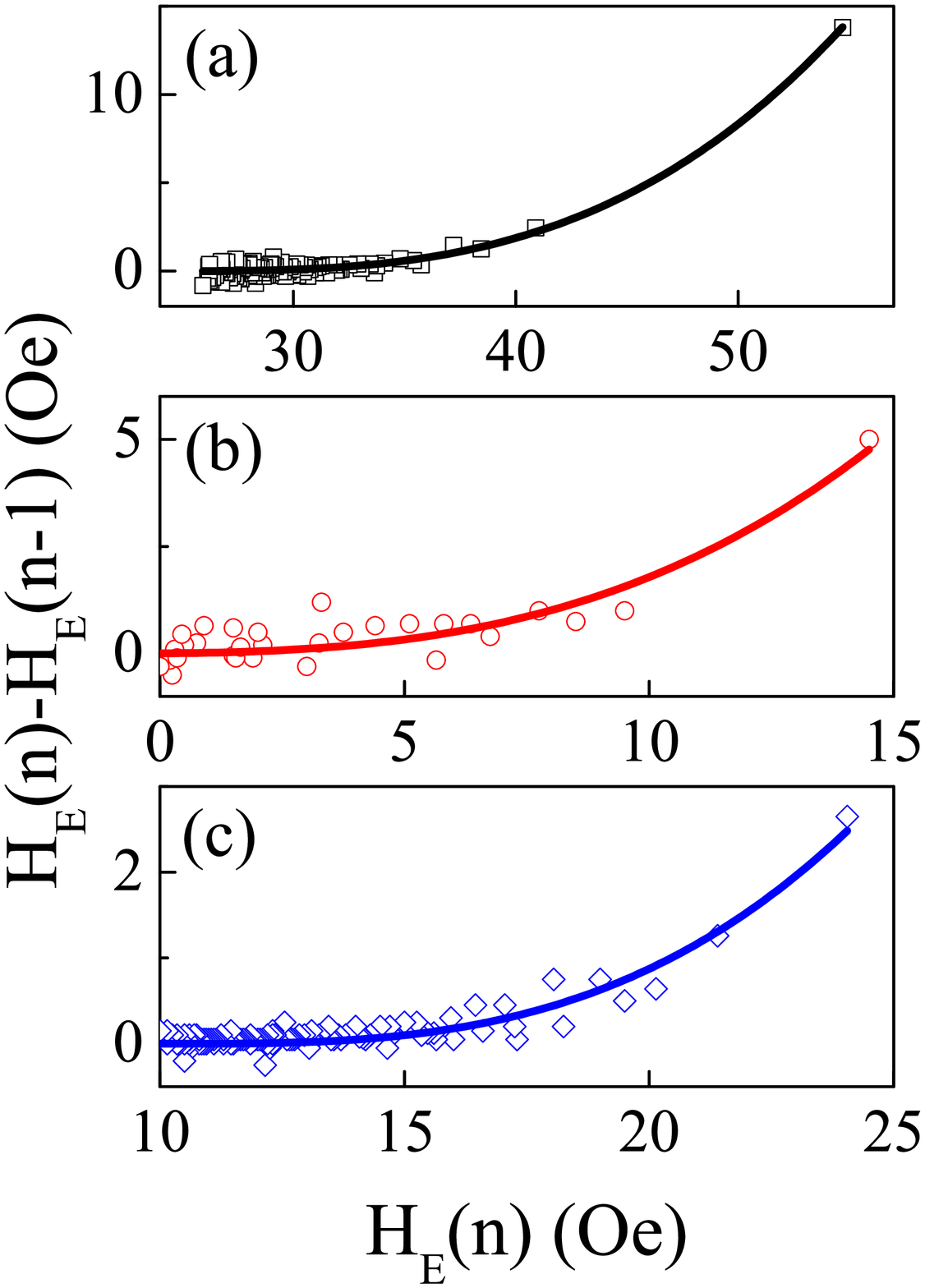}}
\caption{}\label{Fig2}
\end{center}
\end{figure}

\begin{figure}[p]
\begin{center}
\resizebox*{6 in}{!}{\includegraphics*{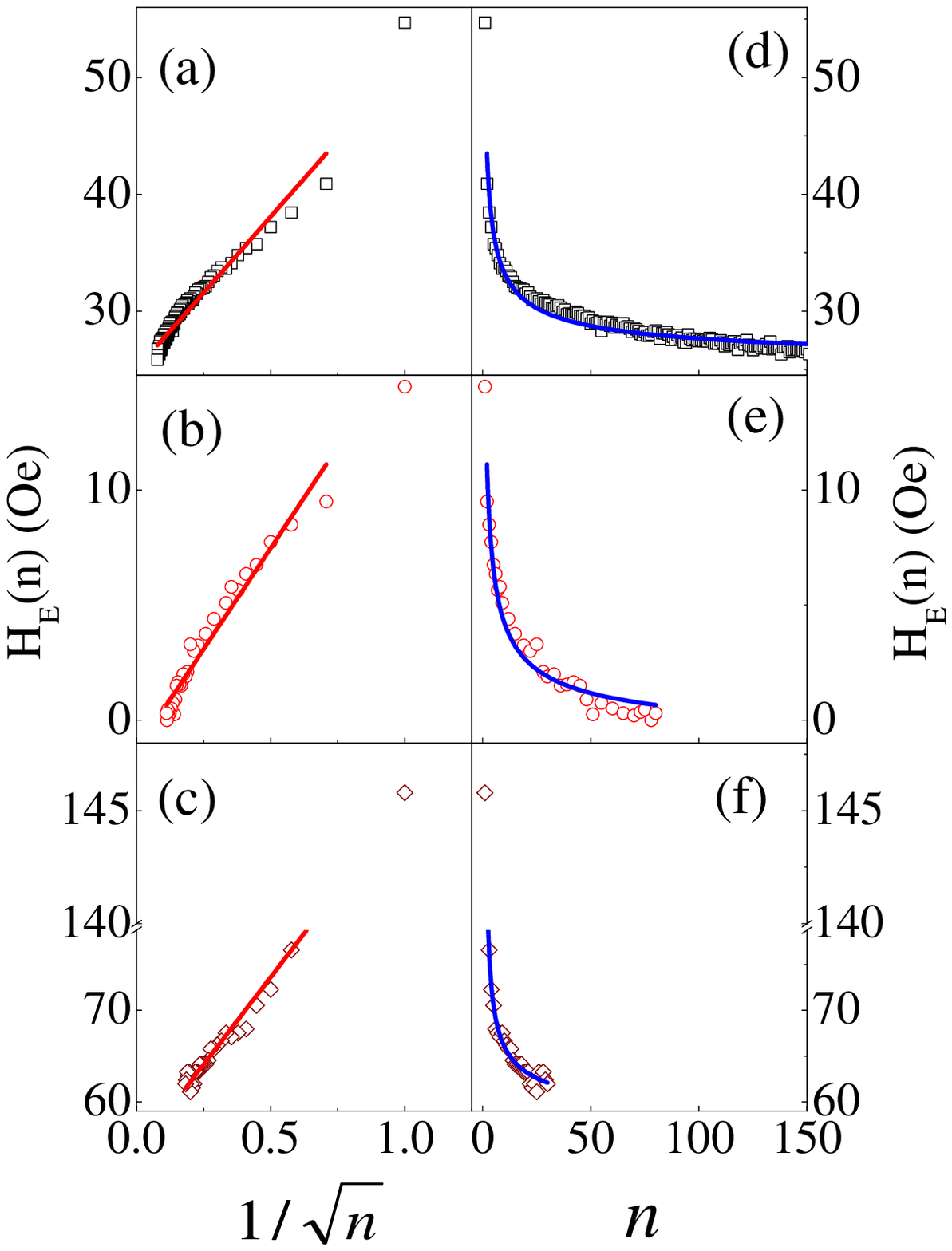}}
\caption{}\label{Fig3}
\end{center}
\end{figure}

\begin{figure}[p]
\begin{center}
\resizebox*{6 in}{!}{\includegraphics*{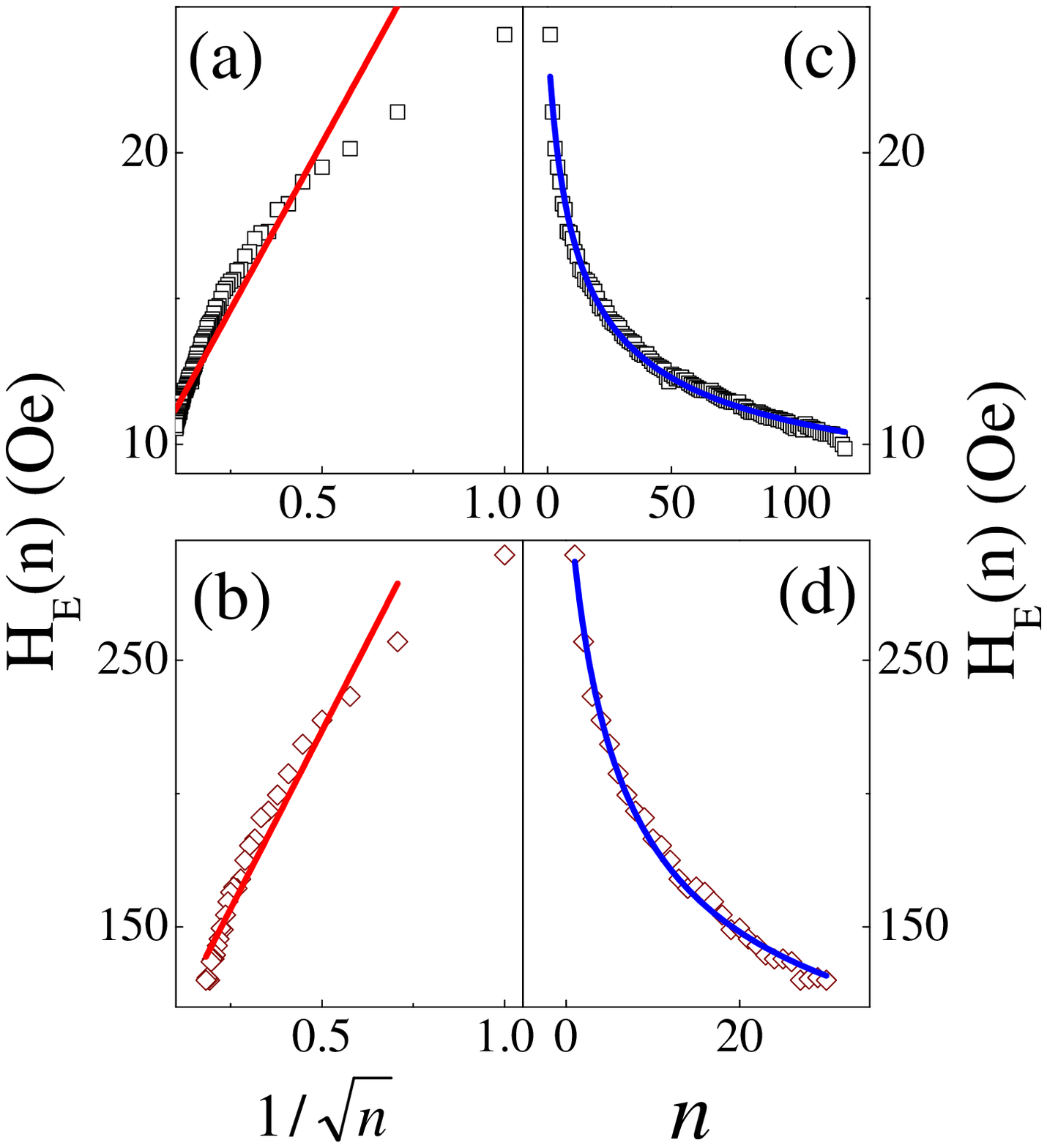}}
\caption{}\label{Fig4}
\end{center}
\end{figure}

\end{document}